\documentclass[%
 reprint,
 amsmath,amssymb,
 aps,
]{revtex4-2}

\usepackage{graphicx}
\usepackage{dcolumn}
\usepackage{bm}

\usepackage{color}

\begin{document}

\preprint{APS/123-QED}

\title{Imperfect bifurcation in the rotation of a propeller-shaped camphor rotor}

\author{Yuki Koyano}
\email{koyano@cmpt.phys.tohoku.ac.jp}
\affiliation{Department of Physics, Graduate School of Science, Tohoku University, Sendai 980-8578, Japan
}

\author{Hiroyuki Kitahata}
 \email{kitahata@chiba-u.jp}
\affiliation{Department of Physics, Graduate School of Science, Chiba University, Chiba 263-8522, Japan
}

\date{\today}

\begin{abstract}
We investigated the bifurcation structure on the self-propelled motion of a camphor rotor at a water surface.
The center of the camphor rotor was fixed by the axis, and it showed rotational motion around it.
Due to the chiral asymmetry of its shape, the absolute values of the angular velocities in clockwise and counterclockwise directions were different.
This asymmetry in the angular velocities implies an imperfect bifurcation.
From the numerical simulation results, we discuss the condition for the occurrence of the imperfect bifurcation.
\end{abstract}

\maketitle

\section{Introduction}

Recently, many kinds of self-propulsion systems have been reported~\cite{Ohta, Ramaswamy2010, Marchetti2013, Bechinger2016, Pimienta2, Michelin2013}.
There are two types of the way for the emergence of self-propulsion; One way is the destabilization of the rest state owing to the spontaneous symmetry breaking.
The other way to induce the self-propulsion originates from the intrinsic asymmetry, which determines the direction of the motion.
For example, a symmetric self-propelled droplet shows translational motion triggered by chemical or thermal fluctuations~\cite{Sumino2005, Domingues, Nagai, Toyota, Izri, Tanaka, Bouillant}, while a self-propelled particle with an anterior-posterior asymmetry shows translational motion in a predetermined direction~\cite{Howse, Lagzi, Jin, Dey, Kang}.
The mechanism of the motion of a symmetric self-propelled particle can be formulated by the bifurcation theory in dynamical systems.
Several bifurcation structures in experimental self-propulsion systems have been reported~\cite{Suematsu2010, Iida2010, NagayamaPhysD2004, Nishi, Koyano2016, Koyano2019, Matsuda}.
It is known that when a slight asymmetry is introduced into a symmetric dynamical system, its bifurcation structure becomes asymmetric, and is called an imperfect bifurcation~\cite{Strogatz}.
An imperfect bifurcation can occur in a self-propelled system~\cite{Ikeda2018}, but it has not been reported yet in experimental systems.

When the center of a self-propelled particle is fixed, it can rotate.
Here, we call such a self-propelled particle a ``rotor''.
A rotor with chiral symmetry can rotate through a spontaneous symmetry breaking~\cite{Pimienta, Takabatake, Bassik, Koyano2017, rotor2, Morohashi}.
On the other hand, the one with a chiral asymmetry rotates in a preferred direction~\cite{NakataLangmuir, Kummel, Frenkel, Frenkel2, Mitsumata, Hayakawa}.
We adopt the motion of such a rotor to investigate the effect of the symmetric property on the self-propulsion, because we can obtain sufficient experimental data without considering the finite system size.

\begin{figure}
	\centering
	\includegraphics{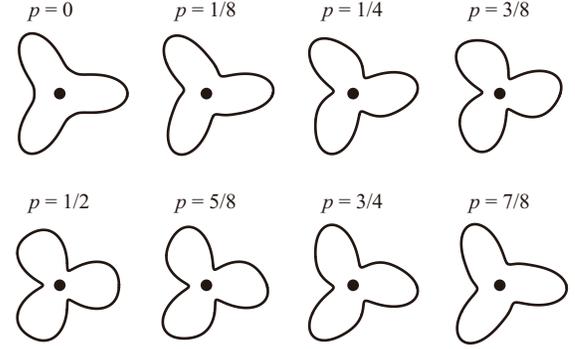}
	\caption{Representative shapes of the propeller-shaped camphor rotors. The mathematical expressions of them are shown in Eq.~\eqref{shape}. Here, we set $a_3 = 0.5$ and $a_6 = 0.1$. The black points represent the centers.}
	\label{fig1}
\end{figure}

As the rotor system where the chiral asymmetry is continuously introduced into a symmetric shape, we consider a propeller-shaped rotor, whose shape is described as
\begin{align}
r =& f(\theta) \nonumber \\
=& R \left[ 1 + a_3 \cos 3 \theta + a_6 \cos 6 \left(\theta - \frac{\pi p}{3} \right) \right] \label{shape}
\end{align}
in two-dimensional polar coordinates $(r,\theta)$.
Several examples of the shape are shown in Fig.~\ref{fig1}.
The parameter $p$ takes a value in $[0,1)$, and corresponds to the chirality of the rotor;
The rotor has no chiral asymmetry for $p = 0$ and $0.5$.
Between $p = 0$ and $0.5$, the chirality of the rotor continuously changes.
Since the shapes of the particles for $p$ and $1-p$ are in mirror symmetry, the magnitudes of chiralities of the rotors are the same though the signs of them are opposite.

To experimentally realize the rotor whose shape is described in Eq.~\eqref{shape}, we adopted a camphor-water system since the shape of the camphor is easily controlled.
The camphor-water system is one of the typical examples of self-propulsion systems, where a camphor particle exhibits the spontaneous motion at a water surface~\cite{Skey1878, Tomlinson1862, Rayleigh1889, NakataLangmuir, Nakata2015, BookChap2}.
The driving force of the camphor particle originates from the surface tension; When the particle is floated at a water surface, it releases camphor molecules around.
Then, the camphor molecules reduce the surface tension of water~\cite{Fujinami1, Fujinami2, SuematsuLang2014}.
It is reported that a symmetric camphor disk moves in a certain direction through spontaneous symmetry breaking~\cite{Hayashima, NagayamaPhysD2004, Chen, Koyano2016, Koyano2019}, while a camphor boat, a plastic plate attached with a camphor disk at its rear, moves in a direction determined by the configuration of the boat~\cite{Kohira, Shimokawa, ikeda_PRE}.
Camphor can be soaked into gel~\cite{Soh} or filter paper~\cite{IkuraPRE} using solvents such as methanol or ethanol.
Arbitrary shapes can be cut out from the gel sheet or filter paper, and thus the shape of the camphor particle can be easily designed.
We made camphor rotors with both the chirally symmetric and asymmetric shapes as described in Eq.~\eqref{shape} using filter paper, and investigated the transition between the motion through a spontaneous symmetry breaking and that in a predetermined direction.

In the present paper, we investigated the rotational motion of a propeller-shaped camphor rotor.
It can rotate both clockwise (CW) and counterclockwise (CCW), but it cannot exhibit translational motion because its center is fixed.
The chirality of the rotor shape was parameterized, and we experimentally investigated the effect of the chirality on the rotational motion of the rotor.
The angular velocities in CW and CCW directions were different when the camphor rotor was chirally asymmetric.
This result implies the imperfect bifurcation.
To investigate the detailed structure of the imperfect bifurcation, we numerically studied the dynamics of the camphor rotor using a mathematical model, in which the rotational motion is coupled with the dynamics of the concentration field of the camphor molecule at a water surface.

\section{Experiments\label{sec_exp}}

\begin{figure}
	\centering
	\includegraphics{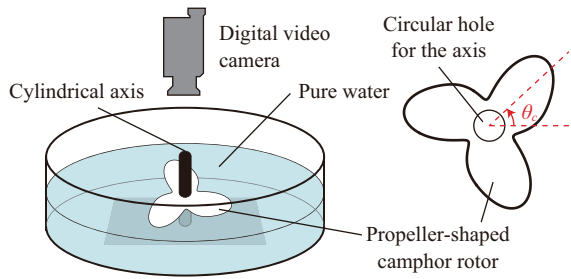}
	\caption{Schematic illustration of the experimental setup. The propeller-shaped rotor has a circular hole at the center for the cylindrical axis.}
	\label{fig2}
\end{figure}

Filter paper (Whatman 1440-240, GE Healthcare Life Science, UK) was cut out into a propeller shape with a hole at its center by a cutting machine (ScaNCut CM300, Brother, Japan).
The propeller-shaped filter paper was soaked into 3~M camphor methanol solution, which was prepared with camphor and methanol purchased from Wako, Japan.
Then it was dried for 600~s in the atmosphere, during which the methanol in filter paper mostly evaporated but camphor remained~\cite{IkuraPRE, EiPhysD2018}.
Then, it was floated at a pure water surface (volume: 750~mL) in a Petri dish (radius: 230~mm), as shown in Fig.~\ref{fig2}.
The pure water was prepared with the Millipore water purifying system (UV3, Merck, Germany).
The center of the propeller-shaped camphor rotor was fixed by the cylindrical axis (radius: 2.5~mm) made with a 3D printer (UP! Plus2, OPT Technologies, Japan).
The behavior of the rotor was recorded with a digital video camera (IVIS HV30, Canon, Japan) from above. All the experiments were carried out at room temperature.

The configuration of the propeller-shaped camphor rotor was described in Eq.~\eqref{shape}.
Here we set $a_3 = 0.5$ and $a_6 = 0.1$ and adopted $p = n/8$ $(n = 0, \cdots, 7)$ (see Fig.~\ref{fig1}).
The mean radius of the camphor rotor $R$ was set to be $R = 10$~mm.
The radius of the hole at the center of the camphor rotor was 2.75~mm so that the camphor rotor could rotate freely but not change the position of its center.

\begin{figure}
	\centering
	\includegraphics{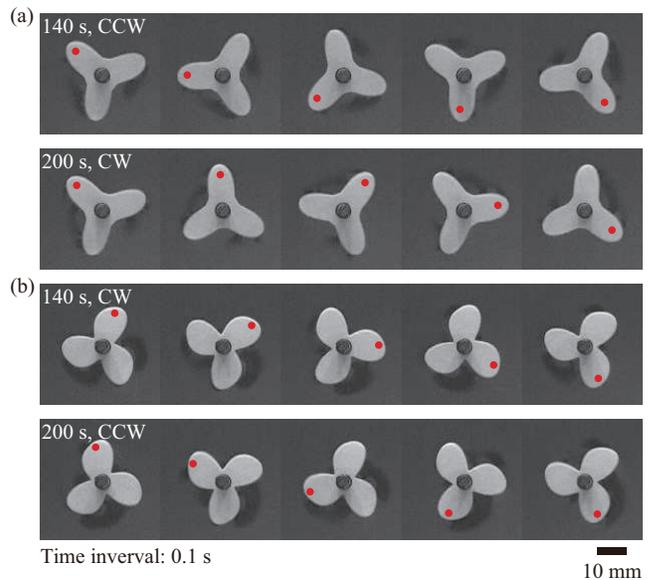}
	\caption{Sequent snapshots of the propeller-shaped camphor rotor, which was rotating in CW or CCW direction. The parameter $p$ was set to be (a) 0 and (b) 0.625. The red (gray) dots were added to the snapshots to indicate the same position on the rotor. The times in the left panels correspond to those in Fig.~\ref{fig4}.}
	\label{fig3}
\end{figure}

\begin{figure}
	\centering
	\includegraphics{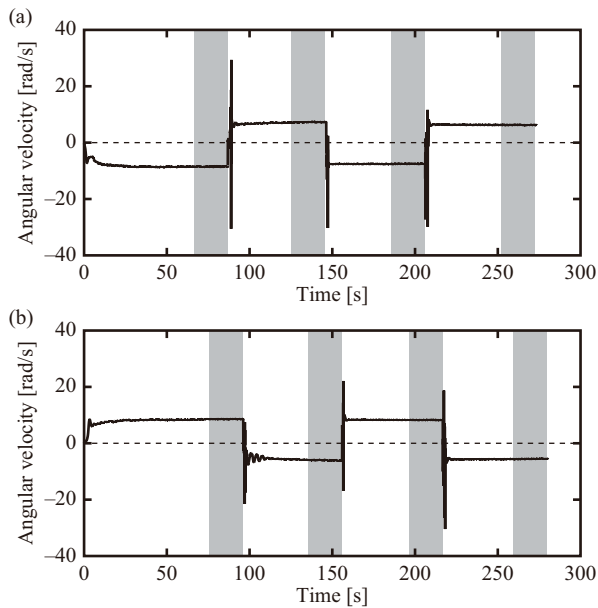}
	\caption{Time series of the angular velocities of the propeller-shaped camphor rotor. The parameter $p$ was set to be (a) 0 and (b) 0.625. The data correspond to the snapshots in Fig.~\ref{fig3}.
	The data in the gray regions (20~s intervals) were used for the calculation of the stationary angular velocities in Fig.~\ref{fig5}.}
	\label{fig4}
\end{figure}

\begin{figure}
	\centering
	\includegraphics{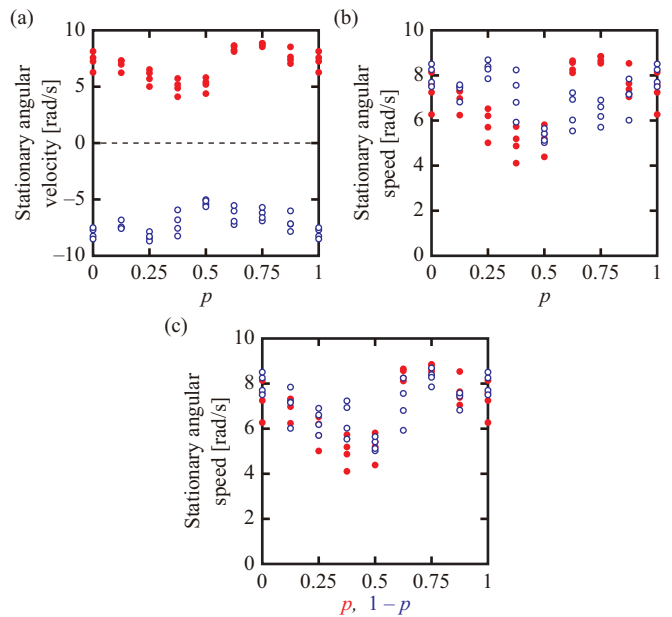}
	\caption{(a) Stationary angular velocities of the propeller-shaped camphor rotor against $p$. (b) Stationary angular speeds against $p$. (c) Stationary angular speeds for the CW direction against $p$ and those for the CCW direction against $1-p$. Red filled circles and blue open circles correspond to CW and CCW rotations, respectively. The angular velocity was averaged for 20 s before the camphor rotor changed its rotational direction. The data at $p = 1$ are identical with those at $p = 0$.}
	\label{fig5}
\end{figure}

When the camphor rotor was floated at a water surface calmly, it began to rotate spontaneously.
The rotation continued at least for 600 s.
To observe the angular velocities for both the CW and CCW rotations, the rotational direction was changed by poking the rotor manually with tweezers after about 90~s from the start of the rotation.
Such operation was repeated every ca.~60~s.
The snapshots of the camphor rotor are shown in Fig.~\ref{fig3}.
The characteristic angle of the rotor $\theta_c$, which is illustrated in Fig.~\ref{fig2}, was obtained by the image processing with ImageJ (NIH)~\cite{IJ}.
The detailed procedure of image processing is shown in Appendix~\ref{A}.
The time evolution of the angular velocities $d \theta_c/dt$ is shown in Fig.~\ref{fig4}.
The angular velocity was relaxed to the terminal value in ca.~10~s after the particle started to rotate, and thus the angular velocity was averaged for 20~s before changing the rotational direction to obtain the stationary angular velocity.
The experiments were repeated twice for each $p$.
The stationary angular velocities for each $p$ are plotted in Fig.~\ref{fig5}.
When the camphor rotor was chirally symmetric, i.e., the cases for $p = 0$ and $0.5$, the stationary angular speed in CW and CCW rotations were almost the same.
In contrast, when the rotor was chirally asymmetric, the angular velocities in CW and CCW rotations were different.
As seen in Fig.~\ref{fig5}(c), the angular speed in the CW direction as the function of $p$ seem to be the same as that in the CCW direction as the function of $1-p$, which is natural from the viewpoint of the symmetry.

\section{Model\label{Sec:model}}

We adopt the following mathematical model describing the dynamics in the present experimental system based on the previous studies~\cite{NagayamaPhysD2004, BookChap2}.
The camphor surface concentration $u$ and the rotational motion of the propeller-shaped camphor rotor characterized by an angle $\theta_c$ are considered in a two-dimensional system.

The time evolution equation of the camphor surface concentration $u(\bm{r},t)$ is described based on the dynamics of the diffusion, evaporation and supply from the rotor, which is described as:
\begin{align}
\frac{\partial u}{\partial t} = D \nabla^2 u - a u + S(\bm{r}; u, \theta_c), \label{eq_conc}
\end{align}
where $D$ is the diffusion coefficient, and $a$ is the evaporation rate of the camphor molecules.
It should be noted that we regard $D$ as the ``effective diffusion coefficient'' including the transport by the Marangoni flow that is induced by the surface tension gradient~\cite{SuematsuLang2014, KitahataJCP2018}.
The function $S(\bm{r}; u, \theta_c)$ represents the supply from the camphor rotor, which is described as:
\begin{align}
S(\bm{r}; u, \theta_c) =& b \Theta(\bm{r}; \theta_c) \left( U_0 - u(\bm{r},t) \right). \label{source}
\end{align}
Here, $b$ is related to the characteristic time scale of the dissolution of camphor molecules, and $U_0$ is the saturated concentration.
$\Theta(\bm{r};\theta_c)$ is a smoothed step function described as
\begin{align}
\Theta(\bm{r}; \theta_c) = \frac{1}{2} \left[1 + \tanh\left( - \frac{d(\bm{r}; \theta_c)}{\delta} \right) \right], \label{step_function}
\end{align}
where $\delta$ is a positive small parameter corresponding to the smoothing length.
The function $d(\bm{r},\theta_c)$ is the signed distance from the periphery of the camphor rotor, which is defined as
\begin{align}
d(\bm{r}; \theta_c) = \left\{
\begin{array}{ll}
\displaystyle{ \min_{0 \leq \phi < 2\pi } \left | \bm{r} - f(\phi-\theta_c) \bm{e}_r(\phi) \right |}, & \bm{r} \notin \Omega(\theta_c), \\
\displaystyle{-\min_{0 \leq \phi < 2\pi } \left | \bm{r} - f(\phi-\theta_c) \bm{e}_r(\phi) \right |}, & \bm{r} \in \Omega(\theta_c).
\end{array} \right.
\end{align}
Here, $\Omega(\theta_c)$ denotes the region covered with the propeller-shaped camphor rotor characterized by the angle $\theta_c$:
\begin{align}
\Omega(\theta_c) = \left \{ \bm{r} = r \bm{e}_r (\theta) \middle | r \leq f(\theta - \theta_c) \right \}.
\end{align}

The dynamics of the rotational motion of the camphor rotor is described as
\begin{align}
I \frac{d^2\theta_c}{dt^2} = - \eta_r \frac{d\theta_c}{dt} + N, \label{eq_thetac}
\end{align}
where $I$ is the moment of inertia of the camphor rotor, $\eta_r$ is the friction coefficient on the rotational motion, and $N$ is the torque originating from the surface tension~\cite{BookChap2, KitahataJPSJ}.
The torque is explicitly calculated as~\cite{KitahataJPSJ}
\begin{align}
N =& \int_{\partial \Omega(\theta_c)} \bm{r}' \times \gamma(u(\bm{r}')) \bm{e}_n(\bm{r}') d\ell' \label{torque1} \\
=& \int_{\Omega(\theta_c)} \bm{r}' \times \nabla_{\bm{r}'} \gamma(u(\bm{r}')) d\bm{r}'. \label{torque2}
\end{align}
Here, $\gamma(u)$ is a function that describes the relation between the surface tension and camphor surface concentration, $\partial \Omega(\theta_c)$ is the periphery of the region $\Omega(\theta_c)$, $\nabla_{\bm{r}'}$ is the vector differential operator with respect to $\bm{r}'$, $\bm{e}_n(\bm{r}')$ is a normal unit vector of the periphery of the particle at $\bm{r}'$, and $d\ell'$ is the arc element of $\partial \Omega(\theta_c)$.
For simplicity, we assume a linear relation between the concentration $u$ and the surface tension $\gamma(u)$ as
\begin{align}
\gamma(u) = \gamma_0 - \Gamma u, \label{surface_tension}
\end{align}
where $\gamma_0$ is the surface tension of pure water, and $\Gamma$ is a positive coefficient.

\section{Numerical calculation\label{sec_NC}}

We performed numerical calculations on the rotation of a propeller-shaped camphor rotor based on the mathematical model.
We introduced the co-rotating frame with an angular velocity $\omega = d\theta_c/dt$, where the supply area of the camphor molecules does not move.
The time evolution of the concentration field is described as
\begin{align}
\frac{\partial u}{\partial t} = \omega \frac{\partial u}{\partial \theta} + D\nabla^2 u - a u + S(\bm{r}; u, 0). \label{adv_dif_eq}
\end{align}
Equation~\eqref{eq_thetac} can be described with $\omega$ instead of $\theta_c$ as:
\begin{align}
I \frac{d \omega}{dt} = - \eta_r \omega + N.
\end{align}
Thus, the time evolutions of $u$ and $\omega$ were numerically calculated.

The parameters of the rotor shape in Eq.~\eqref{shape} were set as $a_3 = 0.5$ and $a_6 = 0.1$, which are the same as those in the experimental setup.
The other parameters were $D = 1$, $a = 0.1$, $b = 10$, $U_0 = 1/A$, $\Gamma = 1$, $I = 10^{-4}$, $\delta = 0.1$, and $R = 1$.
Here, $A$ is the area of the camphor rotor.
The validity of the parameter choices is shown in Appendix~\ref{B}.
The rotor was set at the center of the system.
The concentration field in a circular region with a radius of 10 was considered, which was sufficiently large compared with the rotor size.
The Neumann boundary condition was adopted at the region boundary.
The mesh size was set to be $\Delta x = 0.1$, and we adopted an explicit method for time development with the time step of $\Delta t = 10^{-4}$.
As the initial condition, we set $\omega = \pm 1$, and the concentration field $u(\bm{r},0)$ to be zero.
First, we fixed $\omega$ and only calculated the time evolution of $u$ for $0 \leq t < 1$ in order to stabilize the rotation in the given initial direction.
Then, the time evolution of $\omega$ was calculated together with the time evolution of $u$.
When $|d \omega/dt|$ became less than $10^{-5}$ and the maximum value of $|\partial u/\partial t|$ became less than $10^{-6}$, the angular velocity was regarded to be saturated and was defined as the stationary angular velocity $\omega_f$.

\begin{figure}
	\centering
	\includegraphics{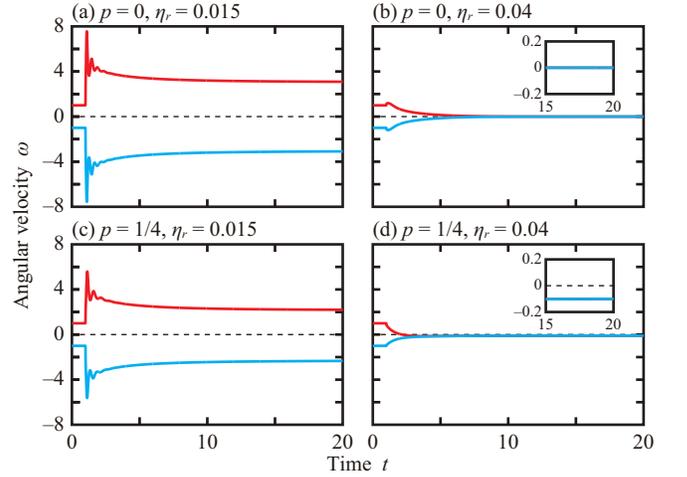}
	\caption{Time series of the angular velocity $\omega$. The parameters $(p, \eta_r)$ are set to be (a) (0, 0.015), (b) (0, 0.04), (c) (1/4, 0.015), and (d) (1/4, 0.04).
	The insets in (b) and (d) show the expanded plots for the final states.
	The angular velocity was fixed as $\omega = 1$ (red) and $\omega = -1$ (cyan) for $0 \leq t < 1$.}
	\label{fig6}
\end{figure}

\begin{figure}
	\centering
	\includegraphics{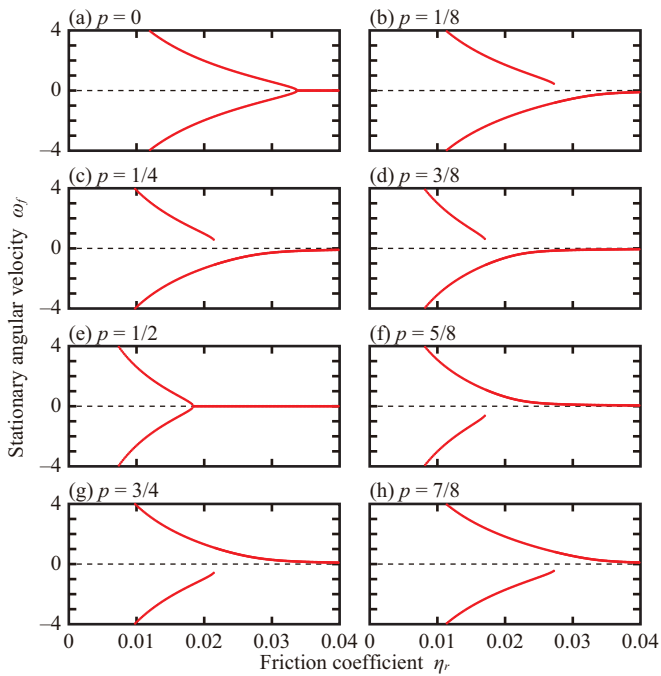}
	\caption{Bifurcation diagram on the stationary angular velocity $\omega_f$ versus the friction coefficient $\eta_r$. The parameter $p$ was set to be (a) $0$, (b) $1/8$, (c) $1/4$, (d) $3/8$, (e) $1/2$, (f) $5/8$, (g) $3/4$, and (h) $7/8$. The results for the two initial conditions $\omega = \pm 1$ are simultaneously plotted.}
	\label{fig7}
\end{figure}

\begin{figure}
	\centering
	\includegraphics{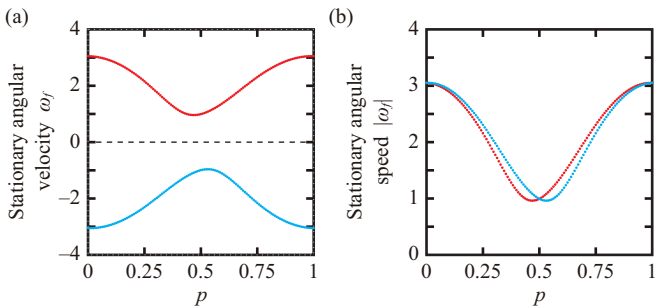}
	\caption{(a) Stationary angular velocities $\omega_f$ versus $p$. (b) Stationary angular speed $\left| \omega_f \right|$ for each $p$. Red and cyan plots correspond to the CCW (positive) and CW (negative) rotations, respectively. $\eta_r$ was set to be $\eta_r = 0.015$.}
	\label{fig8}
\end{figure}

Figure \ref{fig6} shows the time series of the angular velocity $\omega$ for the symmetric ($p=0$) and asymmetric ($p=1/4$) camphor rotors.
In the case of the symmetric camphor rotor, the angular velocity $\omega$ converged to zero for the larger friction coefficient $\eta_r$, while it converged to the finite values with different signs depending on the initial angular velocity for the smaller $\eta_r$.
The stationary angular speeds were the same for both the rotational directions.
In contrast, in the case of the asymmetric camphor rotor, the angular velocity converged to a non-zero value even for the larger $\eta_r$.
For the smaller $\eta_r$, the angular velocity converged to the finite values with different signs depending on the initial rotational direction.
The stationary angular speeds were different from each other.

Figure~\ref{fig7} shows the bifurcation diagram between $\eta_r$ and $\omega_f$.
In the case of the symmetric camphor rotor ($p = 0$ and $p = 1/2$), the supercritical pitchfork bifurcation was observed.
In contrast, the imperfection of the supercritical pitchfork bifurcation was observed for the asymmetric propeller-shaped camphor rotors.
For the larger friction coefficient, the camphor rotor could rotate only in one direction with a small angular velocity.
Here, we call the rotational direction as a preferable direction.
For the smaller friction coefficient, the camphor rotor could rotate in both directions, but the angular speed for the rotation in the preferable direction was greater than that in the unpreferable direction.
By increasing the friction coefficient $\eta_r$, the branch for the rotational motion in the unpreferable direction became unstable, where the saddle-node bifurcation was expected to occur.

The stationary angular velocity $\omega_f$ is also plotted against $p$ in Fig.~\ref{fig8}.
Here, the friction coefficient $\eta_r$ is fixed so that the rotations in both directions are stable.
These results well reproduce the experimental results shown in Fig.~\ref{fig5}.

\begin{figure}
	\centering
	\includegraphics{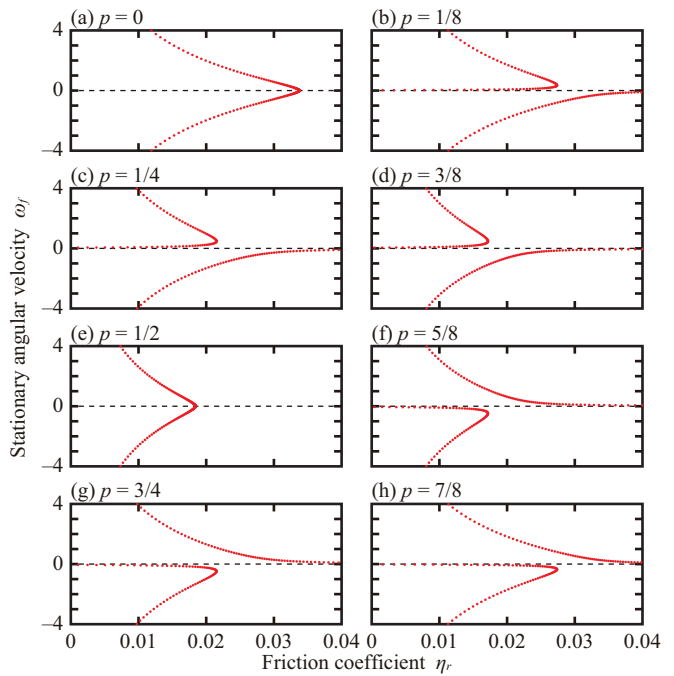}
	\caption{Bifurcation diagram by calculating the steady state for a given fixed angular velocity $\omega$.
	The parameter $p$ was set to be (a) $0$, (b) $1/8$, (c) $1/4$, (d) $3/8$, (e) $1/2$, (f) $5/8$, (g) $3/4$, and (h) $7/8$.
	The bifurcation diagram includes both stable and unstable steady state solutions. The branch of the solution $\omega = 0$ should exist for $p=0$ and $p=1/2$ owing to the symmetric property though it cannot be displayed with this method.}
	\label{fig9}
\end{figure}

To exemplify the bifurcation structure including the unstable steady states, we calculated the converged concentration field $u$ with a given fixed angular velocity $\omega$.
We defined that the concentration field was converged when the maximum value of $|\partial u/\partial t|$ became less than $10^{-6}$.
For the existence of the converged concentration field as the steady state, the friction coefficient $\eta_r$ should satisfy the following relation:
\begin{align}
\eta_r = \frac{N}{\omega}. \label{eta_omega}
\end{align}
We calculated the torque $N$ from the converged concentration field $u$, and obtained the friction coefficient $\eta_r$ satisfying the relation in Eq.~\eqref{eta_omega}.
By changing the angular velocity $\omega$, the bifurcation diagram between $\eta_r$ and $\omega$ was obtained as shown in Fig.~\ref{fig9}.
In this method, both the stable and unstable steady states can be obtained, and Fig.~\ref{fig9} illustrates the structure of the imperfect bifurcation more clearly than Fig.~\ref{fig7}, especially for the saddle-node bifurcation structure seen in the cases of the chirally asymmetric rotors.
In this method, the stabilities of the solutions cannot be determined.
Considering that only the stable solutions are plotted in Fig.~\ref{fig7}, the stability of the solutions in Fig.~\ref{fig9} can be guessed by comparing Fig.~9 with Fig.~\ref{fig7}.
Note that the branch of the solution $\omega = 0$ should exist for $p=0$ and $p=1/2$ owing to the symmetric property though it cannot be calculated with this method.

\section{Discussion}

\begin{figure}
	\centering
	\includegraphics{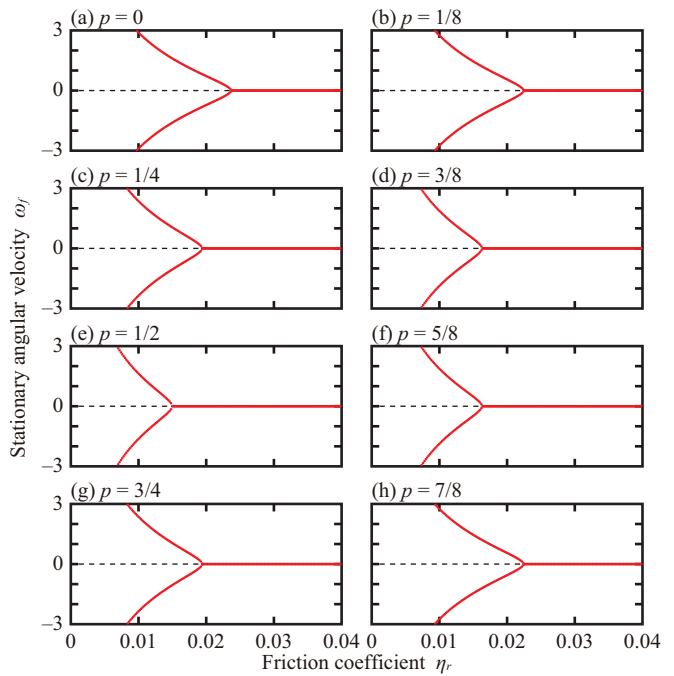}
	\caption{Bifurcation diagram on the stationary angular velocity $\omega_f$ versus the friction coefficient $\eta_r$. 
	The parameter $p$ was set to be (a) $0$, (b) $1/8$, (c) $1/4$, (d) $3/8$, (e) $1/2$, (f) $5/8$, (g) $3/4$, and (h) $7/8$. The results for the two initial conditions $\omega = \pm 1$ are simultaneously plotted.
	This result is comparable with Fig.~\ref{fig7}.
	The difference in the setup between this figure and Fig.~\ref{fig7} is the form of the source term for the concentration field.}
	\label{fig10}
\end{figure}

\begin{figure}
	\centering
	\includegraphics{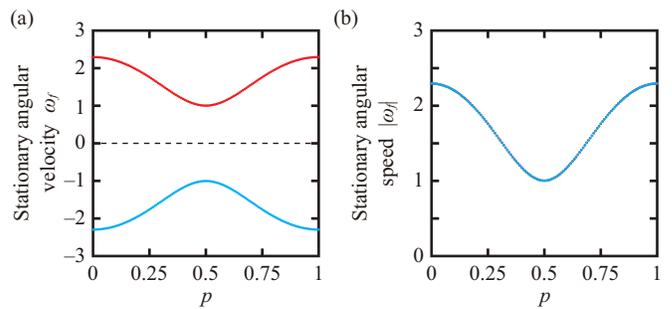}
	\caption{(a) Stationary angular velocity $\omega_f$ for each $p$. (b) Stationary angular speeds $\left| \omega_f \right|$ for each $p$. The plots for both the rotational directions were overlapped.
	Red and cyan plots correspond to the CCW (positive) and CW (negative) rotations, respectively. The friction coefficient $\eta_r$ was set to be $\eta_r = 0.012$.
	This result is comparable with Fig.~\ref{fig8}.
	The difference in the setup between this figure and Fig.~\ref{fig8} is the form of the source term for the concentration field.}
	\label{fig11}
\end{figure}

The source term in Eq.~\eqref{source} for the concentration field represents that the concentration field is saturated at $U_0$.
In the previous studies~\cite{NagayamaPhysD2004, BookChap2}, such saturation was not considered, and more simple source term was adopted:
\begin{align}
\tilde{S}(\bm{r};\theta) =& \frac{S_0}{A} \Theta(\bm{r};\theta_c), \label{source_mod}
\end{align}
where $S_0$ is the total supply per unit time and $A$ is the area of the camphor rotor.
We also attempted the numerical calculations using Eq.~\eqref{source_mod}.
Here, we set $S_0 = 1$, and the other equations and the numerical method were the same as in Secs.~\ref{Sec:model} and \ref{sec_NC}.
We obtained the bifurcation diagram of the stationary angular velocity $\omega_f$ versus the friction coefficient $\eta_r$ as shown in Fig.~\ref{fig10}.
In the current case, the supercritical pitchfork bifurcation was observed for all $p$, which is qualitatively different from Fig.~\ref{fig7}.
The stationary angular velocity $\omega_f$ against $p$ was also obtained as in Fig.~\ref{fig11}.
As seen in Fig.~\ref{fig11}(b), the stationary angular speeds for both rotational directions were the same, which is also qualitatively different from Fig.~\ref{fig8}(b).

The reason why the imperfect bifurcation does not occur can be explained theoretically.
We prove that the torques originating from the surface tension in CW and CCW rotations are the same when we adopt Eq.~\eqref{source_mod} as the source term.
This statement is true even if the rotor shape is asymmetric, and thus, the imperfect bifurcation cannot occur.
The details of the theoretical analysis are shown in Appendix~\ref{C}.

The numerical and theoretical results for the source term in Eq.~\eqref{source_mod} indicate that the supply rate depending on the concentration in Eq.~\eqref{source} is essential for the imperfect bifurcation.
It should be noted that adopting Eq.~\eqref{source} as a source term is just one of the candidates to cause the imperfect bifurcation.
Actually, the imperfect bifurcation occurs when the dependence of the surface tension on the camphor concentration in Eq.~\eqref{surface_tension} is changed to be nonlinear even though the source term in Eq.~\eqref{source_mod} is adopted (data not shown).

We assumed that the Stokes' law-like expression can be adopted for the friction force exerting on a plate on liquid surface, i.e., the friction force is proportional to the velocity of the plate.
Suppose that the friction coefficient per unit area of the plate is $\eta$, the friction force for the rotor can be described as:
\begin{align}
-\int_{\Omega} \bm{r} \times (\eta \bm{v}) \; d\bm{r} 
&= -\eta \dot{\theta}_c \int_{\Omega} r^2 d\bm{r} \\
&\equiv -\eta_r \dot{\theta}_c,
\end{align}
where $\bm{r} = \bm{0}$ is the rotation center and $\Omega$ is the area of the rotor.
Since the rotor shape $\Omega$ is parameterized by $p$, the friction coefficient $\eta_r$ for the rotation changes by 8\% at most depending on the value of $p$.
Such a slight change in the friction coefficient does not affect the bifurcation structure, and thus we neglect the $p$-dependence of $\eta_r$.

It should be noted that Reynolds number in our experiments was $\mathrm{Re} \sim 1000$, and thus, the Stokes' law-like friction may not be good approximation.
Although $\eta_r$ might not be constant but depend on the angular velocity $\dot{\theta}_c$, the bifurcation structure should be conserved as long as $\eta_r (\dot{\theta}_c)$ monotonically changes with regard to $\dot{\theta}_c$.
Moreover, the friction coefficient $\eta_r$ in the CW and CCW rotations can be different for asymmetric rotors.
In our model, the difference of $\eta_r$ in the CW and CCW rotations is neglected.
It can be another candidate to cause the imperfect bifurcation.
More detailed conditions for the occurrence of the imperfect bifurcation should be explored in future work.

\section{Summary}

In this study, we experimentally realized the self-propulsion system in which the imperfect bifurcation occurs.
We designed a propeller-shaped camphor rotor system, where the rotor could only exhibit CW or CCW rotational motion by fixing the center.
By adding the perturbation using tweezers, the propeller-shaped camphor rotor showed the rotations in both directions.
We experimentally confirmed the different stationary angular speeds between CW and CCW directions for the chirally asymmetric rotors, which implies the imperfect bifurcation.
We also investigated the structure of the imperfect bifurcation by numerical calculation.
We numerically reproduced the different stationary angular speeds for CW and CCW rotations in the case of the chirally asymmetric rotors.
By scanning the friction coefficient, the structure of imperfect bifurcation in the stable angular velocities was also revealed.

\begin{acknowledgments}

This work was supported by JSPS KAKENHI Grant Nos.~JP19J00365, JP20K14370, and JP20H02712, and also the Cooperative Research Program of ``Network Joint Research Center for Materials and Devices: Dynamic Alliance for Open Innovation Bridging Human, Environment and Materials'' (Nos.~20201023 and 20204004).
This work was also supported by JSPS and PAN under the Japan-Poland Research Cooperative Program ``Complex spatio-temporal structures emerging from interacting self-propelled particles'' (No.~JPJSBP120204602).

\end{acknowledgments}

\appendix

\section{Detailed procedure on the analyses of the experimental results\label{A}}

In each experiment, we recorded the video of the whitish propeller-shaped camphor rotor from above, and transformed it into the monochronic version. Then we subtracted the background image, which was obtained by detecting the minimum (darkest) value over time at each pixel, and achieved the binarized video in which the regions for the propeller-shaped rotor were extracted. Here we represent the binarized image $\mathcal{F}(x,y)$, which has 1 inside the region of the propeller-shaped rotor, and 0 otherwise, where $(x, y)$ indicates the coordinates of a pixel in the analyzed images.

First, we calculated the center of mass of the propeller-shaped camphor rotor, $(X, Y)$, as follows:
\begin{align}
X = \frac{1}{A} \int x \mathcal{F}(x,y) \, dx dy,
\end{align}
\begin{align}
Y = \frac{1}{A} \int y \mathcal{F}(x,y) \, dx dy,
\end{align}
where $A$ is the area defined as
\begin{align}
A = \int \mathcal{F}(x,y) \, dx dy.
\end{align}
Then, we calculated the coefficients of Fourier series expansion for the 3 mode:
\begin{align}
c_3 =& \frac{1}{A} \int \mathcal{F}(x,y) \cos (3 \theta(x,y;X,Y)) \, dx dy, \\
=& \frac{1}{A} \int \mathcal{F}(x,y) \frac{(x - X)^3 - 3 (x - X) (y - Y)^2}{\left[ (x-X)^2 + (y - Y)^2 \right ]^{3/2}} \, dx dy,
\end{align}
and
\begin{align}
s_3 =& \frac{1}{A} \int \mathcal{F}(x,y) \sin (3 \theta(x,y;X,Y)) \, dx dy, \\
=& \frac{1}{A} \int \mathcal{F}(x,y) \frac{3(x - X)^2(y-Y) - (y - Y)^3}{\left[ (x-X)^2 + (y - Y)^2 \right ]^{3/2}} \, dx dy,
\end{align}
where $\theta(x,y;X,Y)$ is the angle formed by the $x$-axis and the line connecting $(x,y)$ and $(X,Y)$.
Using these values, the characteristic angle $\theta_c$ is calculated as
\begin{align}
\theta_c = \frac{1}{3} \arctan \frac{s_3}{c_3}.
\end{align}

The present procedure is validated as follows:
We set the function representing the shape of the propeller-shaped rotor in the two-dimensional polar coordinates as in Eq.~\eqref{shape}.
Using it, $c_3 A$ and $s_3 A$ are calculated as
\begin{align}
c_3 A = \frac{\pi R^2}{2} a_3 \left(2 \cos 3 \theta_c + a_6 \cos(3 \theta_c + 2 \pi p) \right),
\end{align}
\begin{align}
s_3 A = \frac{\pi R^2}{2} a_3 \left(2 \sin 3 \theta_c + a_6 \sin(3 \theta_c + 2 \pi p) \right).
\end{align}
Then, $\arctan (s_3 / c_3)$ is obtained as
\begin{align}
\arctan \frac{s_3}{c_3} = 3 \theta_c + \arctan \left( \frac{a_6 \sin (2\pi p)}{2 + a_6 \cos(2 \pi p)} \right).
\end{align}
Therefore, $\theta_c$ is obtained as
\begin{align}
\theta_c = \frac{1}{3} \arctan \frac{s_3}{c_3} - \frac{1}{3} \arctan \left( \frac{a_6 \sin (2\pi p)}{2 + a_6 \cos(2 \pi p)} \right).
\end{align}
The second term does not affect the results of the angular velocity $\omega$, since it is constant.

\section{Parameter setting\label{B}}

The units of time, length, and mass are taken as 1 [s], $10^{-2}$ [m], and $10^{-3}$ [kg], respectively.
The diffusion constant $D = 1$ and sublimation rate $a = 0.1$ correspond to $D = 10^{-4} \; \mathrm{[m^2/s]}$ and $0.1 \; \mathrm{[1/s]}$, respectively.
These values are not far from the values obtained by the experiments, $3.94 \times 10^{-3} \; \mathrm{[m^2/s]}$ for the diffusion constant and $1.8 \times 10^{-2} \; \mathrm{[1/s]}$ for the sublimation rate~\cite{SuematsuLang2014}.
As for the value of $b$, we set it arbitrarily ($b=10$) since the experimental quantification of it has not been done.
The radius of the rotor $R = 1$ corresponds to $10^{-2}$ [m], which is consistent with our experiments.

The moment of inertia is set to be $I = 10^{-4}$ so that the friction term is dominant, since the inertia term does not affect the stationary state.
We changed $\eta_r$ in the range of $(0, 0.04]$.
The torque $N$ should be balanced with $\eta_r \omega$ finally, and therefore $N$ is in the order of 0.1.
Using the above scales of time, length, and mass, $N = 0.1$ corresponds to $N = 10^{-8} \; \mathrm{[kg \; m^2/s^2]}$.
This result is consistent with the typical value of the driving force $4.2 \times 10^{-6}$ $\mathrm{[kg \; m/s^2]}$ estimated from the experiments~\cite{SuematsuLang2014}.

\section{Theoretical analysis\label{C}}

In this section, we explain why the imperfect bifurcation does not occur in the case with the source term in Eq.~\eqref{source_mod}.
For simplicity, we consider the sharp shape, i.e., $\delta$ in Eq.~\eqref{step_function} is infinitesimally small, $\delta \to +0$.
In this case, Eq.~\eqref{source_mod} is expressed as follows:
\begin{align}
\tilde{S}(\bm{r};\theta_c) = \left\{ \begin{array}{ll} \dfrac{S_0}{A}, & \bm{r} \in \Omega(\theta_c), \\
0, & \bm{r} \notin \Omega(\theta_c). \end{array} \right. \label{eq_supply}
\end{align}

In the analysis, we use the dimensionless form of the mathematical model, where the spatial, temporal, and concentration scales are chosen as $\sqrt{D/a}$, $1/a$, and $S_0/(a A)$, respectively.
Then, Eqs.~\eqref{eq_conc} and \eqref{eq_supply} are described as
\begin{align}
\frac{\partial u}{\partial t} = \nabla^2 u - u + \tilde{S} (\bm{r}, \theta_c), \label{eq_conc_nd}
\end{align}
\begin{align}
\tilde{S}(\bm{r};\theta_c) = \left\{ \begin{array}{ll} 1, & \bm{r} \in \Omega(\theta_c), \\
0, & \bm{r} \notin \Omega(\theta_c). \end{array} \right.
\end{align}

First, we consider the stationary state of the concentration field.
The solution $u_0$ for the stationary state satisfies the following equation:
\begin{align}
0 = \nabla^2 u_0 (\bm{r}) - u_0 (\bm{r}) + \tilde{S}(\bm{r};\theta_c). \label{ss}
\end{align}
The Green's function $G(\bm{r})$ of Eq.~\eqref{ss} is defined as
\begin{align}
0 = \nabla^2 G(\bm{r}) - G(\bm{r}) + \delta(\bm{r}), \label{def_gf}
\end{align}
which also satisfies
\begin{align}
G(\bm{r}) = G(-\bm{r}). \label{df_prop}
\end{align}
Using the Green's function, the solution $u_0(\bm{r})$ is described as:
\begin{align}
u_0(\bm{r}) = \int_{\Omega(\theta_c)} G(\bm{r} - \bm{r}') d\bm{r}'.
\end{align}

The torque $N_0$ originating from the surface tension depending on $u_0$ is calculated as follows:
\begin{align}
N_0 =& - \Gamma \int_{\Omega(\theta_c)} \bm{r}' \times \nabla' u_0 \; d\bm{r}' \nonumber \\
=& - \Gamma \int_{\Omega(\theta_c)} \int_{\Omega(\theta_c)} \bm{r}' \times \nabla' G(\bm{r}'' - \bm{r}') d\bm{r}'' d\bm{r}' \nonumber \\
=& - \frac{\Gamma}{2} \int_{\Omega(\theta_c)} \int_{\Omega(\theta_c)} (\bm{r}'-\bm{r}'') \times \nabla' G(\bm{r}'' - \bm{r}') d\bm{r}'' d\bm{r}' \nonumber \\
=& 0. \label{N0}
\end{align}
Here, we used Eq.~\eqref{df_prop}, $\nabla' G(\bm{r}'' - \bm{r}') = -\nabla'' G(\bm{r}'' - \bm{r}')$, and that $\nabla' G(\bm{r}'' - \bm{r}')$ is parallel to $(\bm{r}'' - \bm{r}')$.
Thus, if the camphor rotor is stopped, then the torque originating from the surface tension is zero even though the rotor has an asymmetric shape.

When the camphor rotor rotates with a constant angular velocity $\omega$, the concentration field $u$ satisfies the equation:
\begin{align}
-\omega \frac{\partial u(\bm{r})}{\partial \theta} = \nabla^2 u(\bm{r}) - u(\bm{r}) + \tilde{S}(\bm{r};\theta_c). \label{eq_rot}
\end{align}
Here, $u$ is expanded with regard to $\omega$,
\begin{align}
u(\bm{r}) = \sum_{n=0}^\infty \omega^n u_n(\bm{r}). \label{u_exp}
\end{align}
The concentration field of the 0-th order $u_0$ satisfies
\begin{align}
0 = \nabla^2 u_0 - u_0 + \tilde{S}(\bm{r};\theta_c), \label{u0}
\end{align}
whereas the concentration field of the $n$-th order $u_n$ $(n \geq 1, n \in \mathbb{N})$ satisfies
\begin{align}
-\frac{\partial u_{n-1}}{\partial \theta} = \nabla^2 u_n - u_n. \label{un}
\end{align}
The solutions for Eqs.~\eqref{u0} and \eqref{un} are calculated as
\begin{align}
u_0 (\bm{r}) =& \int_{\mathbb{R}^2} G(\bm{r} - \bm{r}') \tilde{S}(\bm{r}';\theta_c) d\bm{r}', \label{u0_s} \\
u_n (\bm{r}) =& \int_{\mathbb{R}^2} \frac{\partial u_{n-1}}{\partial \theta'} G(\bm{r} - \bm{r}') d\bm{r}'. \label{un_s}
\end{align}
Using the expressions in Eqs.~\eqref{u0_s} and \eqref{un_s}, the explicit forms of $u_n$ is obtained as:
\begin{align}
&u_n(\bm{r}) = \int_{\mathbb{R}^2} \cdots \int_{\mathbb{R}^2} G(\bm{r} - \bm{r}_0) \nonumber \\
& \times \prod_{k=1}^n \left [ \frac{\partial}{\partial \theta_k} G(\bm{r}_{k-1} - \bm{r}_k ) \right ] \tilde{S}(\bm{r}_n;\theta_c) d\bm{r}_n \cdots d\bm{r}_0.
\end{align}

Since the concentration field $u$ is expanded with regard to the angular velocity $\omega$, and the torque $N$ in Eq.~\eqref{torque2} is linear for the concentration field $u$, the torque $N$ is also expanded with regard to $\omega$ as follows:
\begin{align}
N = \sum_{n=0}^\infty \omega^n N_n. \label{N_exp}
\end{align}
If the angular velocities are the same for both CW and CCW rotations, the torque $N$ should be an odd function with regard to $\omega$.

The torque $N_0$ originating from the surface tension $\gamma(u_0(\bm{r}))$ has already been calculated in Eq.~\eqref{N0}.
Here, the torque $N_2$ originating from the surface tension $\gamma(\omega^2 u_2(\bm{r}))$ is calculated:
\begin{widetext}
\begin{align}
N_2 =& - \omega^2 \Gamma \int_{\Omega(\theta_c)} \bm{r}' \times \nabla' u_2(\bm{r}') d\bm{r}' \nonumber \\
=& - \omega^2 \Gamma \int_{\Omega(\theta_c)} \bm{r}' \times \nabla' \left ( \int_{\mathbb{R}^2} G(\bm{r}'' - \bm{r}') \frac{\partial}{\partial \theta''} \int_{\mathbb{R}^2} G(\bm{r}''' - \bm{r}'') \frac{\partial}{\partial \theta'''} \int_{\Omega(\theta_c)} G(\bm{r}''''-\bm{r}''') d\bm{r}'''' d\bm{r}''' d\bm{r}'' \right ) d\bm{r}' \nonumber \\
=& - \omega^2 \Gamma \int_{\Omega(\theta_c)} \int_{\mathbb{R}^2} \int_{\mathbb{R}^2} \int_{\Omega(\theta_c)} \bm{r}' \times \nabla' G(\bm{r}'' - \bm{r}') \frac{\partial}{\partial \theta''} G(\bm{r}''' - \bm{r}'') \frac{\partial}{\partial \theta'''} G(\bm{r}''''-\bm{r}''') d\bm{r}'''' d\bm{r}''' d\bm{r}'' d\bm{r}' \nonumber \\
=& - \omega^2 \Gamma \int_{\Omega(\theta_c)} \int_{\mathbb{R}^2} \int_{\mathbb{R}^2} \int_{\Omega(\theta_c)} \frac{\partial}{\partial \theta'} G(\bm{r}'' - \bm{r}') \frac{\partial}{\partial \theta''} G(\bm{r}''' - \bm{r}'') \frac{\partial}{\partial \theta'''} G(\bm{r}''''-\bm{r}''') d\bm{r}'''' d\bm{r}''' d\bm{r}'' d\bm{r}' .
\end{align}
By changing the variables of integration, we have
\begin{align}
N_2 =& - \omega^2 \Gamma \int_{\Omega(\theta_c)} \int_{\mathbb{R}^2} \int_{\mathbb{R}^2} \int_{\Omega(\theta_c)} \frac{\partial}{\partial \theta''''} G(\bm{r}''' - \bm{r}'''') \frac{\partial}{\partial \theta'''} G(\bm{r}'' - \bm{r}''') \frac{\partial}{\partial \theta''} G(\bm{r}'-\bm{r}'') d\bm{r}' d\bm{r}'' d\bm{r}''' d\bm{r}''''.
\end{align}
By using the relations $\partial_{\theta'} G(\bm{r}'' - \bm{r}') = -\partial_{\theta''} G(\bm{r}'' - \bm{r}')$ and $G(\bm{r}'' - \bm{r}') = G(\bm{r}' - \bm{r}'')$, we have
\begin{align}
N_2 =& \omega^2 \Gamma \int_{\Omega(\theta_c)} \int_{\mathbb{R}^2} \int_{\mathbb{R}^2} \int_{\Omega(\theta_c)} \frac{\partial}{\partial \theta'''} G(\bm{r}''' - \bm{r}'''') \frac{\partial}{\partial \theta''} G(\bm{r}'' - \bm{r}''') \frac{\partial}{\partial \theta'} G(\bm{r}'-\bm{r}'') d\bm{r}' d\bm{r}'' d\bm{r}''' d\bm{r}'''' \nonumber \\
=& \omega^2 \Gamma \int_{\Omega(\theta_c)} \int_{\mathbb{R}^2} \int_{\mathbb{R}^2} \int_{\Omega(\theta_c)} \frac{\partial}{\partial \theta'''} G(\bm{r}'''' - \bm{r}''') \frac{\partial}{\partial \theta''} G(\bm{r}''' - \bm{r}'') \frac{\partial}{\partial \theta'} G(\bm{r}''-\bm{r}') d\bm{r}' d\bm{r}'' d\bm{r}''' d\bm{r}'''' \nonumber \\
=& -N_2.
\end{align}
\end{widetext}
Thus, the torque proportional to $\omega^2$ should be zero.
Such calculation can be done in a parallel way for the even-number order of $\omega$.
Therefore, if the camphor molecules are supplied constantly in the area of the rotor $\Omega(\theta_c)$, the torques originating from the surface tension should be the same for CW and CCW rotations, even though the shape is chirally asymmetric.
Thus, we should consider other asymmetries in our system to explain the imperfection of the bifurcation structure.

Here, we discuss the physical intuition why the imperfection cannot occur in the model with constant supply of camphor in Eq.~\eqref{eq_supply}.
To simplify the discussion, we consider the force instead of the torque.
Let us consider two points, $P$ and $Q$, inside the camphor rotor region $\Omega$.
The force $F_{Q \to P}$ ($F_{P \to Q}$) exerting on the point $P$ ($Q$) originating from the concentration field by the dissolution of camphor from the point $Q$ ($P$) is considered.
Since the concentration field formed by each point is described by the Green's function, the forces $F_{Q \to P}$ and $F_{P \to Q}$ should be balanced.
Such force balance holds for any pair of points inside $\Omega$.
Therefore, the force originating from the surface tension is zero when a camphor particle with an arbitrary shape stops.
This explanation can be also extended to the torque.
In this section, we have showed mathematically.

The essence of the proof is that the areas supplying the camphor molecules and receiving the force originating from the surface tension are the same.
Thus, in the model with Eq.~\eqref{eq_supply}, a banana-shaped camphor particle can stop.
Such conclusion is not consistent with the experimental results~\cite{NakataLangmuir}, and thus we adopt Eq.~\eqref{source} as a supply term.
It should be noted that a camphor boat cannot stop since the areas supplying the camphor molecules and receiving the force originating from the surface tension are different~\cite{Kohira, Shimokawa, ikeda_PRE}.

\end{document}